# Visualizing Double-Slit Interference on a Shoestring 

Beth Parks ⬤ ; Hans Benze

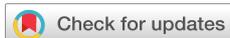 Check for updates

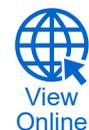
View
Online

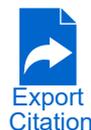
Export
Citation

# Visualizing Double-Slit Interference on a Shoestring

**Beth Parks and Hans Benze,** Colgate University, Hamilton, NY

S tudent misconceptions of the double-slit experiment (Fig. 1) are abundant. The most common ones that we observe include: (1) belief that constructive interference requires both pathlengths to be integer multiples of the wavelength ("$L_1 = n_1\lambda$" and "$L_2 = n_2\lambda$") rather than only the pathlength difference ($\Delta L \equiv |L_1 - L_2| = n\lambda$); (2) failure to understand that the justification for $d \sin\theta = n\lambda$ is that $d \sin\theta$ is a good approximation to $\Delta L$; (3) confusion about the limits in which the approximation $d \sin\theta = n\lambda$ is valid; and (4) confusion about the limits in which the approximation $\sin\theta = \tan\theta = y/D$ is valid. Most of these misconceptions have been observed in the past.[1,2] We consider the first of these to be the most damaging, since it will hinder students from understanding interference in other contexts in the future. To address this misconception, we have designed an exercise that strongly reinforces the correct understanding ($\Delta L = n\lambda$) while also helping overcome the other misconceptions.

For teaching a different interference geometry (Bragg scattering), we previously developed a laboratory exercise that uses wavelengths in the centimeter range (microwaves) to help students visualize the scattering geometry and the condition for constructive interference.[3] Here we describe a similar approach for teaching double-slit interference, increasing the wavelength to the centimeter range, extending a demonstration recently described by Michael Bell.[4]

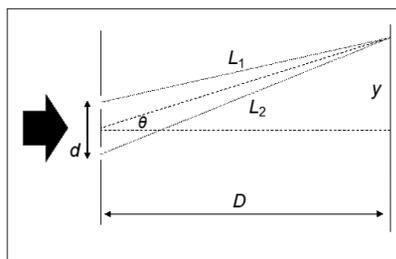

**Fig. 1. Double-slit interference geometry. Light passes through slits spaced by $d$ and hits a screen at distance $D$.**

Bell used strings to illustrate the path of light from the two slits to the observation point and marked the strings with tape flags to represent wavelengths. When the two strings intersect at tape flags, then the two pathlengths differ by an integer number of wavelengths ($\Delta L = n\lambda$), indicating constructive interference.

Bell's demonstration does a nice job of helping students see, on a macroscopic scale, how $\Delta L$ varies with position. However, unless it is extremely carefully presented, it will tend to reinforce the *incorrect* belief that constructive interference occurs when both pathlengths are integer numbers of wavelengths (misconception 1). Also, it doesn't easily lend itself to measuring lengths and checking approximations (misconceptions 2, 3, and 4). For that reason, we have developed a variation that can be used as a demonstration or—as we prefer—expanded into a hands-on lab exercise.

In an earlier version of this exercise (Fig. 2), we used tape measures to represent the wave paths and asked students to assume that the wavelength was 10 cm. They recorded values of $y$ (as defined in Fig. 1) where the two waves would interfere constructively. While this method was somewhat successful, students needed to be guided carefully to understand that they were looking for points where the two pathlengths differed by 10 cm, 20 cm, 30 cm, . . . . Otherwise, students spent much time looking for exactly what matched their misconception: points where both pathlengths were multiples of 10 cm, or where the pathlength had increased by 10 cm from its value at $y = 0$ in the center of the pattern.

Our current apparatus, shown both in a photograph and schematically in Fig. 3, uses 10-foot long rainbow-striped roller-derby shoelaces.[5] The edge of the table in the photograph represents the observation screen, and moving the paperclip that holds the shoelaces allows the students to change the position of the observation point $y$ along this screen. Students match colors and record the values of $y$ at which the two patterns align. Almost any rope or ribbon with a repeating pattern in the range of 3 to 10 cm in length would work, but the waxed shoe laces with $\lambda = 8.83$ cm are particularly convenient to work with because they resist tangling or stretching. Students also measure the two pathlengths $L_1$ and $L_2$ in terms of the wavelength. Since the repeated pattern is subdivided into six color bars, it is easy for students to estimate these distances to fractions of the wavelength. Also, the rainbow pattern is suggestive of the color scheme that is often used to represent phase in a wave, and the bright color scheme makes it quite obvious when the two pathlengths differ by an integer number of wavelengths.

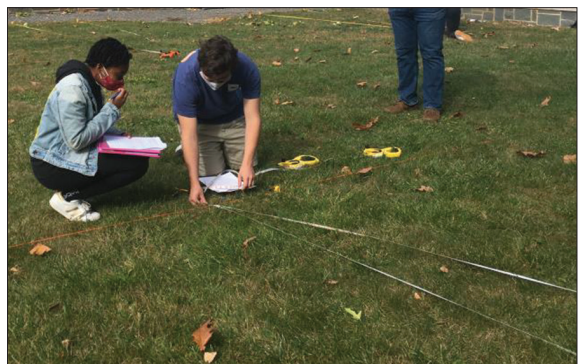

**Fig. 2. In an earlier version of the exercise, students used tape measures to represent the paths and measure their lengths.**

We used a marker to number each wavelength repetition on the laces (black writing on purple squares in Fig. 3), which also had the advantage of making clear which side was "up." (Unfortunately, the stripes on the reverse side are not aligned with those on the front side.) The laces are attached to the wall or the far edge of the table at the top of the red square numbered "0." Students record the distances $y$ and the lengths of each path at the positions where the waves align. For example, for the situation in Fig. 3 where the paper clip is just past the fourth color block (green), students would record that the left path had length $(27 + 4.5/6)\lambda$ and the right path had length $(25 + 4.5/6)\lambda$, making the pathlength difference $\Delta L = 2\lambda$. If the laces are laid across a tabletop, attaching 100-g weights to the ends of the laces creates sufficient tension to eliminate slack. The exercise could also be done with the laces stretched out on the floor and a student providing the tension. In that case, the "screen" could be represented by a line of tape on the floor.

**Table I.** The values of *y* were initially measured in meters and converted to wavelength units, using $\lambda = 0.0883$ m. The values $d \sin \theta = yd/(y^2 + D^2)^{1/2}$ and $yd/D$ were calculated using the measured *y*'s and the measured values of $d = 1.064$ m and $D = 1.826$ m. See the "experimental tips" section for a discussion of measurement.

| *y* (m) | *y*/$\lambda$ | $L_1/\lambda$ | $L_2/\lambda$ | $\Delta L/\lambda$ | $d \sin \theta/\lambda$ | $(yd/D)/\lambda$ |
|---------|---------|---------|---------|---------|---------|---------|
| 0.000 | 0 | 21.51 | 21.51 | 0 | 0.00 | 0.00 |
| 0.155 | 1.76 | 22.07 | 21.07 | 1 | 1.02 | 1.02 |
| 0.319 | 3.61 | 22.75 | 20.75 | 2 | 2.07 | 2.11 |
| 0.489 | 5.54 | 23.63 | 20.63 | 3 | 3.12 | 3.23 |
| 0.668 | 7.57 | 24.67 | 20.67 | 4 | 4.14 | 4.41 |
| 0.863 | 9.77 | 25.96 | 20.96 | 5 | 5.15 | 5.70 |

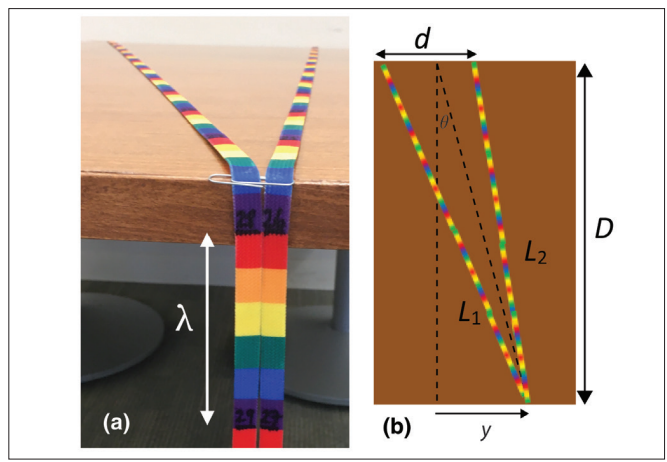

**Fig. 3.** (a) The condition for constructive interference is achieved when the patterns on the laces line up. The number of wavelengths can be seen in black handwriting on purple squares. (b) (Schematic) Top view of the apparatus.

Sample results are shown in Table I. Students can see that the approximation $\Delta L = d \sin \theta$ holds fairly well over the entire range, but the small-angle approximation $\sin \theta = y/D$ breaks down at the larger angles. Instructors may actually be surprised to see how well the approximation $\Delta L = d \sin \theta$ holds even for this case where $d/D = 0.58$; see articles by Poon and Sobel for interesting geometric arguments why this is the case.[6,7]

## Experimental tips:

- $\lambda$ can be found most accurately by measuring the distance *D* both in meters and in wavelengths, then taking the ratio of those two numbers.

- If space is limited, place the "slits" towards one side of the table, so that the whole width of the table is available for measurements to the other side of the center line, enabling measurements at higher $\theta$.

- The measurements of $L/\lambda$ can be more accurate than one might expect, since each block is 1/6 of a wavelength and the position can be estimated to 1/4 of a block.

- The two laces used in this trial had wavelengths that differed by about 1%. To obtain the results in Table I, we applied different tensions to the laces to bring the two wavelengths into agreement: we hung 100 g on one lace and 200 g on the other. Without that correction, measurements of *y* differed by several centimeters from the calculated positions of the maxima. However, even with the differing $\lambda$'s, the data are accurate enough to convince students that, while $\Delta L = d \sin \theta$ is a good approximation over the whole range, $\Delta L = yd/D$ breaks down at higher angles.

This exercise can easily be completed in less than an hour, leaving time for students to also observe the optical double-slit interference pattern using a laser. At this point, it should be clear to students that the approximations $\Delta L = d \sin \theta$ and $\sin \theta = y/D$ both hold extremely well for the optical double-slit experiment, resulting in the observed evenly spaced maxima.

## References

1. Bradley S. Ambrose, Peter S. Shaffer, Richard N. Steinberg, and Lillian C. McDermott, "An investigation of student understanding of single-slit diffraction and double-slit interference," *Am. J. Phys.* **67**, 146–155 (Feb. 1999).

2. Karen Wosilait, Paula R. L. Heron, Peter S. Shaffer, and Lillian C. McDermott, "Addressing student difficulties in applying a wave model to the interference and diffraction of light," *Phys. Educ. Res., Am. J. Phys. Suppl.* **67**, S5–S15 (July 1999).

3. Joseph C. Amato and Roger E. Williams, "Rotating crystal microwave Bragg diffraction apparatus," *Am. J. Phys.* **77**, 942–945 (Oct. 2009).

4. Michael Scott Bell, "Double ring stand interference," *Phys. Teach.* **58**, 204–205 (March 2020).

5. "Rainbow block 120-inch waxed shoe and skate laces," available from https://derbylaces.com, $9.99 per pair.

6. Dick C. H. Poon, "How good is the approximation 'path difference ≈ $d \sin \theta$'?" *Phys. Teach.* **40**, 460–462 (Nov. 1999).

7. Michael I. Sobel, "Algebraic treatment of two-slit interference," *Phys. Teach.* **40**, 402–404 (Oct. 2002).

**Beth Parks** *teaches at Colgate University in Hamilton, NY, and holds the position of visiting professor at Mbarara University of Science and Technology in Mbarara, Uganda, where she served as a Fulbright fellow in 2015-16. She also serves as editor of the* American Journal of Physics. **meparks@colgate.edu**

**Hans Benze** *is the laboratory technician at Colgate University, where he builds and designs lab apparatus and demonstrations.*